\documentclass[journal=jacsat,manuscript=article]{achemso}

\usepackage{chemformula} % Formula subscripts using \ch{}
\usepackage[T1]{fontenc} % Use modern font encodings

\author{Xuyang Chang}
\affiliation[BIT]
{School of Information and Electronics \& Advanced Research Institute of Multidisciplinary Sciences, Beijing Institute of Technology, Beijing 100081, China}
\author{Shaowei Jiang}
\affiliation[Uconn]{Department of Biomedical Engineering, University of Connecticut, Storrs, Connecticut 06269, USA}
\author{Yongcun Hu}
\affiliation[BIT]
{School of Information and Electronics \& Advanced Research Institute of Multidisciplinary Sciences, Beijing Institute of Technology, Beijing 100081, China}

\author{Liheng Bian}
\affiliation[BIT]
{School of Information and Electronics \& Advanced Research Institute of Multidisciplinary Sciences, Beijing Institute of Technology, Beijing 100081, China}
\email{bian@bit.edu.cn}
\title{Pixel super-resolved lensless on-chip sensor with scattering multiplexing}

\begin{document}

\begin{tocentry}
\begin{center}
\includegraphics[width=\linewidth]{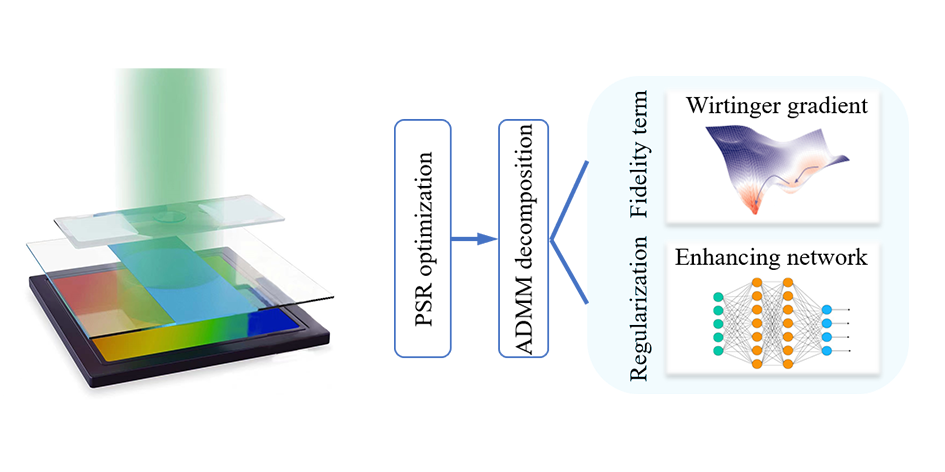}
\end{center}
The integrated lensless on-chip sensor and the pixel super-resolution technique.
\end{tocentry}

\begin{abstract}
Lensless on-chip microscopy has shown great potential for biomedical imaging due to its large-area and high-throughput imaging capabilities. By combining the pixel super-resolution (PSR) technique, it can improve the resolution beyond the limit of the imaging detector. However, existing PSR techniques are restricted to the feature size and crosstalk of modulation components (such as spatial light modulator), which cannot efficiently encode target information. Besides, the reconstruction algorithms suffer from the trade-off between image quality, reconstruction resolution and computational efficiency. In this work, we constructed a novel integrated lensless on-chip sensor via scattering multiplexing, and reported a robust PSR algorithm for sample reconstruction. The sensor employed a scattering layer as a modulator, which was permanently integrated with the detector. Benefiting from the high-degree-of-freedom reconstruction of the scattering layer, we realized fine wavefront modulation with a small feature size. The integration engineering avoided repetitious calibration and reduce the measurement complexity. The reported PSR algorithm combines both model-driven and data-driven strategies to efficiently exploit the high-frequency information from the fine modulation. A series of experiments validated that the reported sensor provides a low-cost solution for large-scale microscopic imaging, with significant advantages in resolution, image contrast and noise robustness.
\end{abstract}

\section{Introduction}
High-resolution imaging across a wide field of view (FOV) is essential for cell morphological explorations and resolving underlying mechanisms of cellular activity \cite{fan2019video,park2021review}. However, the development of microscopy has long been confounded by the trade-off between the imaging resolution and FOV, which restricts the space-bandwidth product and information throughput \cite{zheng2013wide}. Recent advanced computational imaging techniques provide effective solutions, such as Fourier ptychographic microscopy \cite{zheng2013wide,zheng2021concept}, synthetic aperture microscopy \cite{tippie2011high,gao2022resolution}, whole slide imaging \cite{ghaznavi2013digital,bian2020autofocusing,jiang2022high}, and lensless on-chip microscopy \cite{greenbaum2012imaging,ozcan2016lensless}. Among these methods, the lensless on-chip microscopy is an attractive technique due to its lens-free, low-cost and digitalized characteristics, which has been widely used in high-throughput screening and artificial intelligence diagnosis \cite{seo2009lensfree,stybayeva2010lensfree,mudanyali2010compact,jiang2022blood,jiang2022ptychographic}. 

The resolution of lensless on-chip microscopy is mainly restrained by the pixel size of the detector \cite{zhang2017adaptive}. Nevertheless, the array detector may fail to meet the Nyquist criteria \cite{shannon1949communication} due to the discrete sampling. Pixel super-resolution (PSR) techniques \cite{gao2020high,gao2021generalized,chang2022plug} can be employed to achieve a resolution beyond the pixel size limit of the detector. The essence of PSR is to increase the observation diversity by means like sub-pixel shifting \cite{zheng2011epetri}, multiple wavelengths \cite{luo2016pixel}, illumination angles or distances \cite{zhang2017adaptive} and wavefront modulation \cite{gao2021generalized,jiang2021resolution,chang2022plug}. A series of low-resolution (LR) intensity-only measurements are acquired to digitally synthesize a high-resolution (HR) complex wavefront.

Among the above-mentioned methods, the wavefront modulation method is one generalized modality for various lensless on-chip holographic imaging. Numerous studies have revealed the relations between the imaging parameters (such as the wavelength, the distance between the sample to the detector, the modulation pattern profile, sample type and the pixel size of the detector) and imaging quality/resolution \cite{katkovnik2017computational,zhang2020resolution}. The modulation feature size is also a critical factor, but relevant research is not available, to the best of our knowledge. The root cause is the size limit of modulation elements employed in the imaging set-ups. For example, most spatial light modulators (SLM) are manufactured using liquid crystal materials and they suffer from crosstalk between adjacent pixels, resulting in errors in the modulation process \cite{moser2019model,gao2019self}. The common consequence is the smoothing effect of the modulation patterns, which expands the modulation size. Thus, the final effective modulation feature size is larger than the detector’s pixel size, leading to the coarse modulation of the wavefront.

In this work, we report a novel lensless on-chip sensor via scattering multiplexing, together with a joint optimization PSR reconstruction algorithm. Inspired by recent studies the disordered scattering media can be utilized as 'lenses' to encode the high-frequency information \cite{vellekoop2010exploiting,mosk2012controlling,lee2016exploiting}. We employed a thin scattering layer for wavefront modulation and integrated it with the image detector. Then, we shift the integrated sensor to different $x-y$ positions to capture multiple diffraction images. These measurements and reconstruction algorithm was utilized to calibrate the scattering layer and recover the HR sample, without any additional components and setups. 
%placed a scattering layer between the sample and the detector and moved it to different $x-y$ positions to modulate the wavefront. A series of intensity-only measurements which correspond to different positions were captured to recover a high-resolution wavefront. 
Benefiting from the high-degree-of-freedom reconstruction of the scattering layer, we can obtain a smaller modulation feature size compared with the detector's pixel size. The fine wavefront modulation can efficiently encode target information, improving both the imaging quality and resolution. To exploit the high-frequency information during the fine modulation and against measurement noise, a robust PSR reconstruction algorithm is reported. It employs the alternating direction method of multipliers (ADMM) \cite{boyd2011distributed} framework to decompose the complex PSR reconstruction problem into several subproblems. Then, we introduced the Wirtinger gradient \cite{kreutz2009complex,gao2021generalized} and enhancing neural network \cite{zhang2018ffdnet} to tackle the data fidelity and regularization terms. In this way, the reported reconstruction algorithm enjoys the advantages of both model-driven and data-driven, and brings strong robustness, high fidelity, and wide generalization for PSR reconstruction.

\section{Methods}

\subsection{The reported scattering multiplexing lensless on-chip sensor}

\begin{figure}[t!]
\centering
\includegraphics[width=\linewidth]{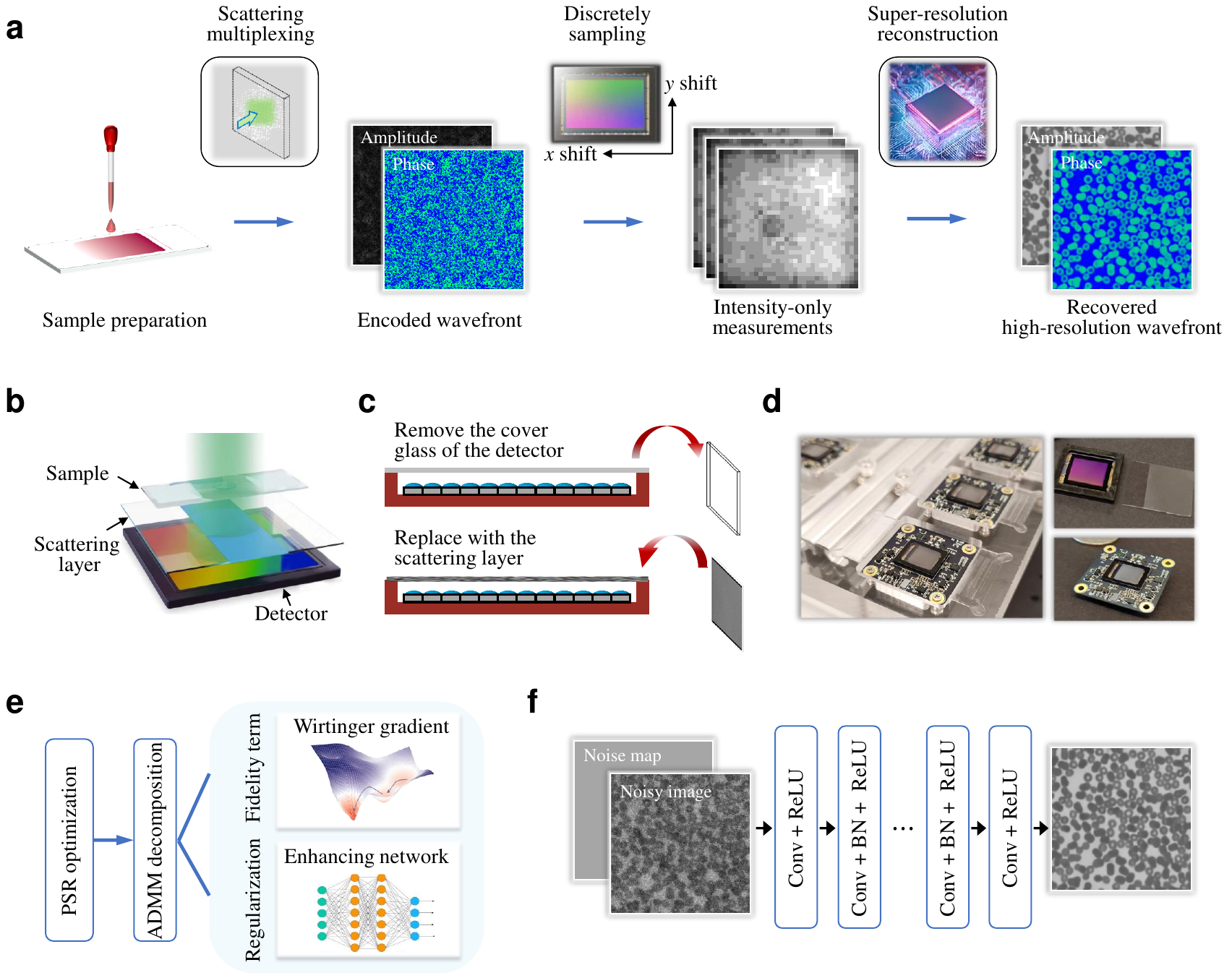}
\caption{Architecture of the reported sensor and the reconstruction algorithm. (a) presents its workflow. The wavefront of the sample is encoded by the scattering layer and captured by the detector. Then, multiple measurements are used to recover a high-resolution wavefront of the sample. (b) - (d) are the integration engineering. The scattering layer is permanently attached to the sensor to avoid repetitious calibration. (e) is the framework of the reported robust PSR reconstruction algorithm. We introduce the Wirtinger gradient and the enhancing neural network to tackle the data fidelity and regularization term. (f) shows the architecture of the enhancing neural network. It contains several combinations of convolution (Conv), batch norm (BN) and rectified linear unit (ReLU).}
\label{Fig1_system_architect}
\end{figure}

\textbf{Workflow.} Figure \ref{Fig1_system_architect} (a) presents the workflow of the reported scattering multiplexing lensless on-chip sensor. Overall, the sample's wavefront propagates to the scattering layer plane and is encoded by a random scattering medium. The scattering layer brings two advantages. First, it can convert the large diffraction angles of sample exit waves into smaller angles, which can facilitate the PSR reconstruction beyond the limitation imposed by the system transfer function. Second, the modulation pattern of the scattering layer can be calibrated freely and its feature size is smaller than the detector's pixel size, realizing fine wavefront modulation and promoting PSR reconstruction. After modulation, the encoded wavefront propagates to the sensor plane. Finally, multiple intensity-only measurements are used to reconstruct a high-resolution sample's wavefront. 

Mathematically, the forward model of the reported sensor can be indicated as
%\begin{equation}
%I_{j}=\left|\left[\left(u(x, y) * \mathcal{P}\left(d_{1}\right)\right) \cdot D\left(x-x_{j}, y-y_{j}\right)\right] * \mathcal{P}\left(d_{2}\right)\right|^{2},
%\label{Eq1}
%\end{equation}
\begin{equation}
\left\{\begin{array}{c}
\mathcal{W}=u\left(x, y\right) * \mathcal{P}\left(d_{1}\right) \\
I_{j}=\left|\left[\mathcal{W}_{j}\left(x+x_{j}, y-y_{j}\right) \odot D\right] * \mathcal{P}\left(d_{2}\right)\right|^{2}
\end{array}\right.
\label{Eq1}
\end{equation}
where $(x, y)$ is the spatial coordinates, $u(x, y)$ is the exit wavefront at the sample plane, $\mathcal{W}_{j}\left(x+x_{j}, y-y_{j}\right)$ is the wavefront of sample at the scattering layer plane which corresponds to $(x_{j},y_{j})$ positional shift, $D$ is the scattering layer's complex profile, $I_{j}$ is the $j$th ($j$ = 1,2,$\dots$,$J$) intensity-only measurements at the detector plane, $'*'$ denotes the convolution operation and $\mathcal{P}(d)$ represents the free space propagation over a distance of $d$, which was described by the Rayleigh–Sommerfeld diffraction model
\begin{equation}
\label{Eq2}
\mathcal{P}(d)=\mathcal{F}^{-1}\left\{H\left(f_{x}, f_{y}, d\right) \cdot \mathcal{F}\left[\mathcal{W}  \right]\right\},
\end{equation}
where $(f_{x}, f_{y})$ represents the frequency coordinates, $\mathcal{W}$ is the target wavefront, $\mathcal{F}$ and $\mathcal{F}^{-1}$ are 2D Fourier transform (FT) and inverse FT, respectively. $H\left(f_{x}, f_{y}, d\right)$ is the transfer function which follows the angular spectrum theory \cite{steward2004fourier} as
\begin{equation}
\label{Eq3}
H= \left\{\begin{array}{cl}
\exp \left[i \frac{2 \pi}{\lambda} z \sqrt{1-\lambda^{2}\left(f_{x}^{2}+f_{y}^{2}\right)}\right], & f_{x}^{2}+f_{y}^{2} \leq \frac{1}{\lambda^{2}} \\
0 & \text { otherwise },
\end{array}\right.
\end{equation}
where $\lambda$ represents the wavelength. 

Considering discrete sampling and subsequent PSR reconstruction, a brief formula of the forward model can be summarized as
\begin{equation}
\label{Eq4}
\textbf{I}=\textbf{U}\left|\textbf{A} \textbf{u}\right|^{2}+\omega,
\end{equation}
where $\textbf{A}$ is the above scattering multiplexing and propagation process. $\textbf{U}$ represents undersampling due to discrete detection. $\omega$ is the measurement noise.\\ %Assuming that the detector pixel size is $\theta_{D} \times \theta_{D}$, and that of the high-resolution target is $\theta_{C} \times \theta_{C}$, the following introduced PSR reconstruction engages to solve the problem of $\theta_{D}>\theta_{C}$.
\textbf{Integration engineering.} Figure \ref{Fig1_system_architect} (b) - (d) show the reported sensor's diagram and the integration method. We permanently attached the scattering layer to the detector. Specifically, the cover glass of the detector was heated by a hot air gun and then removed. After that, we applied nail polish to the edge of the bare detector for bonding the scattering layer. The integrated sensor requires calibration before imaging (Supplementary Note 1). Once characterized, the integrated sensor can realize super-resolution coherent imaging with joint amplitude and phase information.

\subsection{The reported robust PSR reconstruction algorithm}
As indicated in Eq. (\ref{Eq4}), the reported robust PSR algorithm solves the undersampling problem and retrieves the phase information. Figure \ref{Fig1_system_architect} (e) presents the reported PSR algorithm. It employs the ADMM framework, by introducing an auxiliary parameter $v$, the general PSR reconstruction can be expressed as
\begin{equation}
\label{Eq5}
(\hat{\boldsymbol{u}}, \hat{\boldsymbol{v}})=\underset{\boldsymbol{u}, \boldsymbol{v}}{\operatorname{argmin}} f(\boldsymbol{u})+\eta g(\boldsymbol{v}), \text { subject to } \boldsymbol{u}=\boldsymbol{v}
\end{equation}
where $f(u)$ is the data-fidelity term which can be regarded as the loss of the forward model, $g(u)$ is the regularization term and $\eta$ is the weight coefficient.

Equation (\ref{Eq5}) can be decomposed into the following three subproblems \cite{yuan2020plug}
\begin{equation}
\begin{aligned}
\label{Eq6}
&\boldsymbol{u}^{(k+1)}=\operatorname{argmin}_{\boldsymbol{u}} f(\boldsymbol{u})+\frac{\rho}{2}\left\|\boldsymbol{u}-\left(\boldsymbol{v}^{(k)}-\frac{1}{\rho} \boldsymbol{t}^{(k)}\right)\right\|_{2}^{2} \\
&\boldsymbol{v}^{(k+1)}=\operatorname{argmin}_{\boldsymbol{v}} \eta g(\boldsymbol{v})+\frac{\rho}{2}\left\|\boldsymbol{v}-\left(\boldsymbol{u}^{(k)}+\frac{1}{\rho} \boldsymbol{t}^{(k)}\right)\right\|_{2}^{2} \\
&\boldsymbol{t}^{(k+1)}=\boldsymbol{t}^{(k)}+\rho\left(\boldsymbol{u}^{(k+1)}-\boldsymbol{v}^{(k+1)}\right),
\end{aligned}
\end{equation}
where the superscript $(k)$ is the iteration number. Then, we introduced the Wirtinger gradient \cite{kreutz2009complex,gao2021generalized} and denoising network \cite{zhang2018ffdnet} to tackle the first two subproblems.\\
\textbf{Solving $\boldsymbol{u}$:} the update of $\boldsymbol{u}$ is based on gradient descent
\begin{equation}
\boldsymbol{u}^{(k+1)}=\boldsymbol{u}^{(k)}-\gamma^{(k)} \nabla f\left(\boldsymbol{u}^{(k)}\right),
\label{Eq7}
\end{equation}
where $\gamma$ is the step size in a range from 0.5 to 2, and $\nabla f$ represents the gradient. As suggested that amplitude residual $L=\sqrt{|\boldsymbol{A} \boldsymbol{u}|^{2}}-\sqrt{\boldsymbol{I}}$ has better performance than the intensity one \cite{yeh2015experimental,gao2021generalized}. Thus, we employed the Wirtinger gradient \cite{kreutz2009complex} to calculate the complex-valued gradient of the amplitude residual (Supplement Note 2), which can be expressed as 
\begin{equation}
\nabla f(\boldsymbol{u})=\left(\frac{\partial f(\boldsymbol{u})}{\partial \boldsymbol{u}}\right)^{*}=\boldsymbol{A}^{*} \operatorname{diag}(\boldsymbol{A} \boldsymbol{u}) \operatorname{diag}\left(\frac{1}{\sqrt{|\boldsymbol{A} \boldsymbol{u}|^{2}}}\right) L.
\label{Eq8}
\end{equation}\\
\textbf{Solving $\boldsymbol{v}$:} the update of $\boldsymbol{v}$ is using the denoising algorithm directly
\begin{equation}
\label{Eq9}
\boldsymbol{v}^{(k+1)}=\mathcal{D}\left(\boldsymbol{u}^{(k)}+\frac{1}{\rho} \boldsymbol{t}^{(k)}\right),
\end{equation}
where $\mathcal{D}$ represents the denoiser. In recent years, convolutional neural network (CNN) shows great advantages in image enhancement tasks, such as denoising, super resolution and deblurring \cite{zhang2017beyond,zhang2018learning,peng2021u}. CNN can learn a mapping from degraded images to clear images directly, providing efficient solutions with low computational complexity and satisfactory performance. We employed the FFDNet \cite{zhang2018ffdnet} as the denoiser due to its state-of-the-art performance \cite{kong2020comprehensive} and flexible setup, as shown in Fig. \ref{Fig1_system_architect} (f). FFDNet contains a noise map as an input parameter to control the denoising degree, which can effectively balance the detail maintenance and smoothness during each iteration.

\section{Results}
\subsection{Simulations.}
We first implemented a series of simulations to validate the advantage of small modulation feature sizes. In our simulations, two natural images were employed as the latent amplitude and phase, as presented in Fig. \ref{Fig3_Noise}. The value range of the amplitude image was [0,1] and the phase image was [0,$\frac{\pi}{2}$]. The wavelength was 532 $nm$, the pixel size of detector were 5.6 $\mu m$ - 11.2 $\mu m$, and the computational pixel size $\Delta_{C}=1.4$ $\mu m$. We set different modulation sizes ($\Delta_{M}=1.4-11.2$ $\mu m$) under different undersampling ratios ($\theta=4, 6, 8$), and studied their effect on the reconstruction quality which was quantified by peak signal to noise ratio (PSNR).       

\begin{figure}[t!]
\centering
\includegraphics[width=\linewidth]{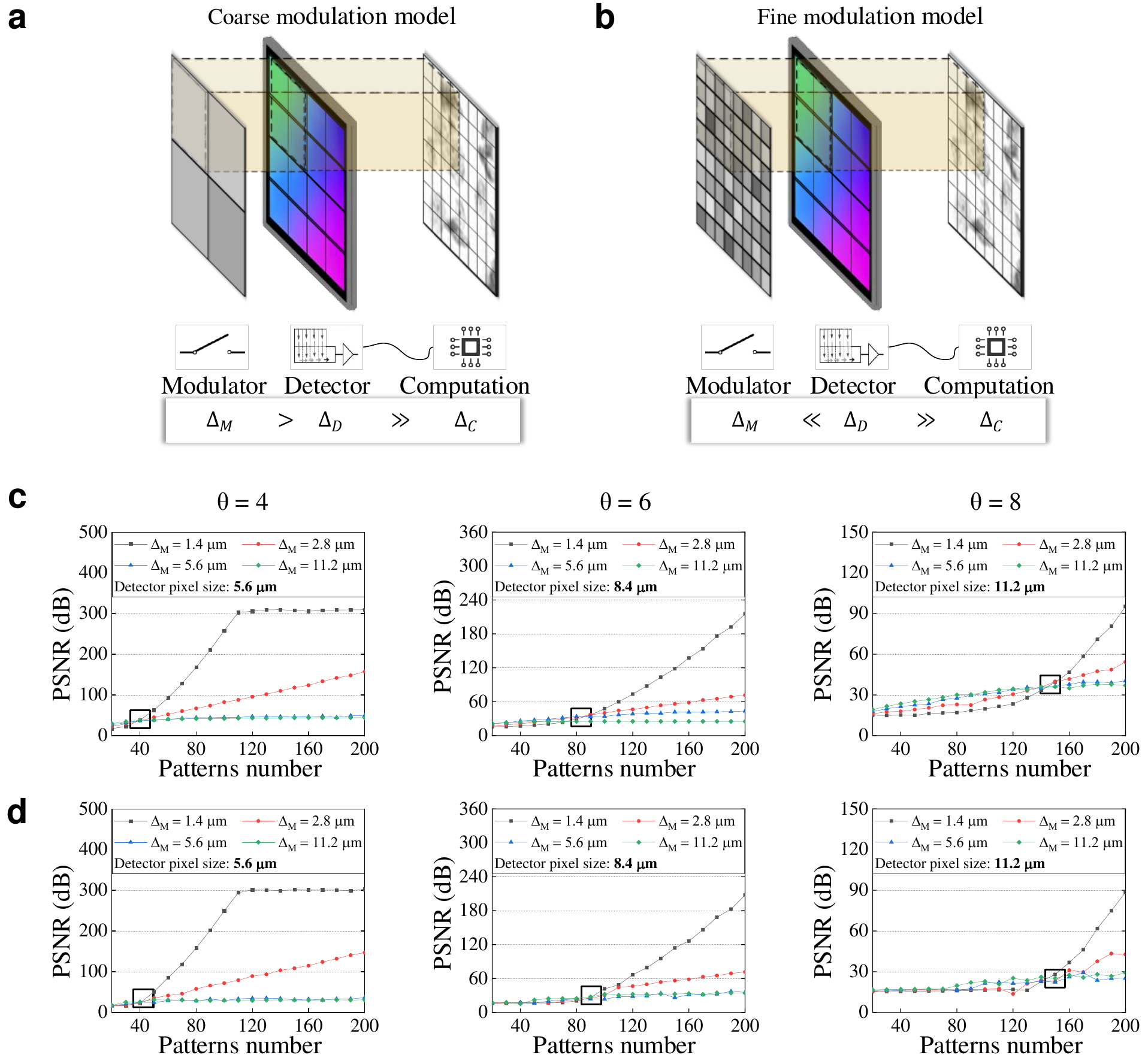}
\caption{Simulation results of different modulation feature sizes. (a) - (b) present the pixel relations in the coarse modulation model and fine modulation model, respectively. To overcome crosstalk, the equivalent pixel size of the modulator is larger than the detector and computation pixel sizes in the conventional coarse modulation model. We employed a scattering layer to replace the conventional modulator, realizing fine wavefront modulation. (c) - (d) are amplitude and phase results of different modulation sizes and undersampling ratios under noiseless measurements. The results show that the smaller modulation sizes conduce to high imaging quality when the pattern number is adequate.}
\label{Fig2_Noiseless}
\end{figure}

Figure \ref{Fig2_Noiseless} (a) shows the coarse and fine modulation models. The modulation feature size is larger than the detector's pixel size in the coarse model and is far smaller than the detector's pixel size in the fine model. Figure \ref{Fig2_Noiseless} (b) presents the quantitative results using the Wirtinger Flow (WF) \cite{gao2021generalized} algorithm under noiseless measurement. The abscissa is the patterns number (measurements number) and the ordinate is the PSNR index. The results validate that the smaller modulation sizes conduce to high imaging quality when the pattern number is adequate. From the aspect of modulation size, with the decrement of $\Delta_{M}$, the advantage of fine modulation is more prominent. Quantitatively, the size of $\Delta_{M}=1.4$ $\mu m$ obtains at most one order of magnitude improvement on PSNR index compared with the size of $\Delta_{M}=11.2$ $\mu m$ (as shown in the phase result under $\theta=4$). From the aspect of patterns number, there is a threshold which is determined by $\theta$ to characterize the measurement redundancy, as indicated in the black rectangular. For higher undersampling ratios, more patterns were required to serve fine modulation.

\begin{figure}[t!]
\centering
\includegraphics[width=\linewidth]{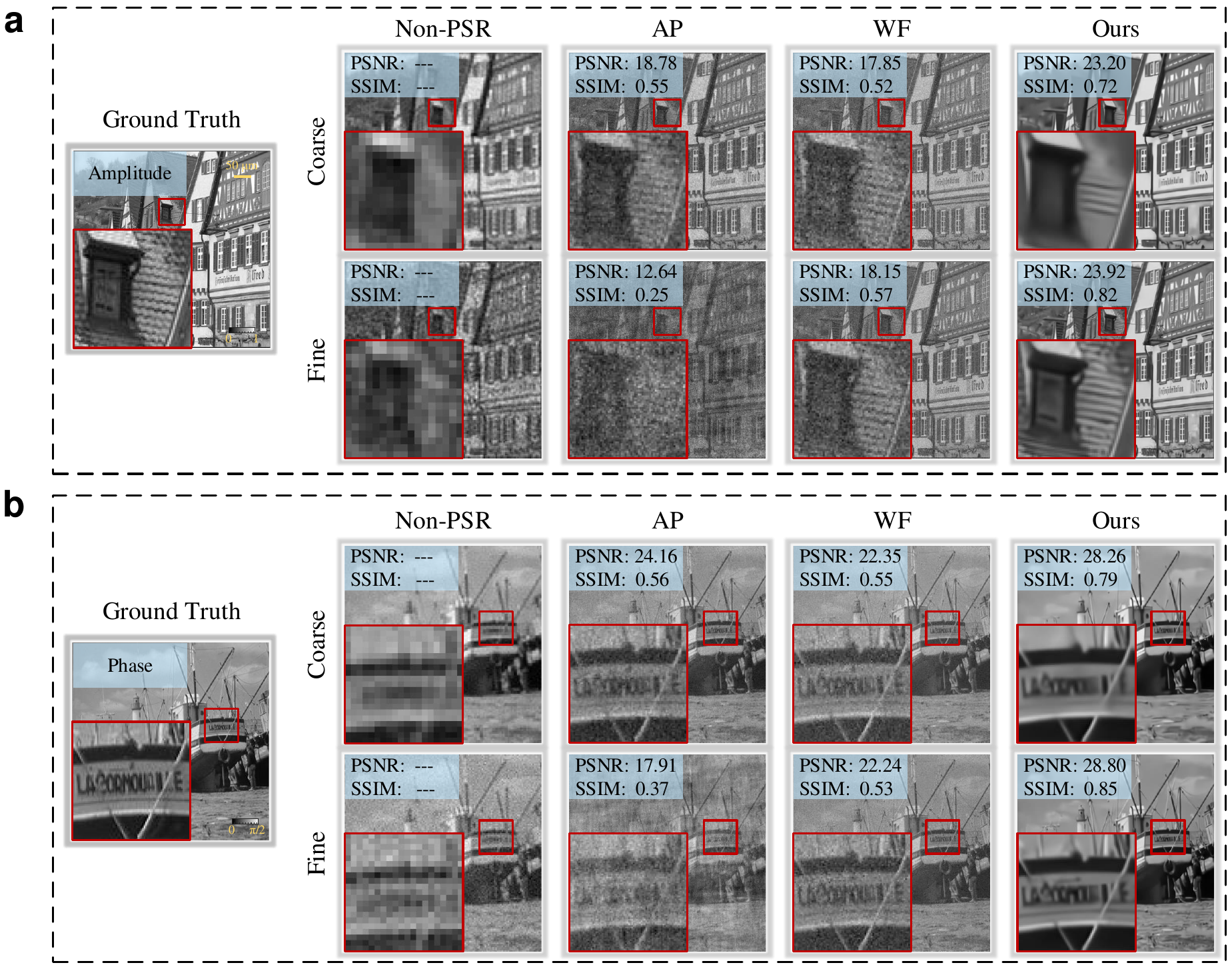}
\caption{Simulation results of different reconstruction algorithms under 5dB measurement noise. (a) and (b) are amplitude and phase results, respectively. The patterns number was 100, $\theta=4$, $\Delta_{M}$ in coarse and fine modulation models were 5.6 $\mu m$ and 1.4 $\mu m$, respectively. }
\label{Fig3_Noise}
\end{figure}

We further studied the impact of measurement noise and the performance of the reported reconstruction algorithm. The comparison methods include alternating projection (AP) \cite{gerchberg1972practical} and WF \cite{gao2021generalized}. Specifically, we added 5 dB (signal-noise ratio) Gaussian noise, and compared these algorithms under coarse ($\Delta_{M}=5.6$ $\mu m$) and fine modulations ($\Delta_{M}=1.4$ $\mu m$). The pattern number here was 100 and the undersampling ratio was 4. Figure \ref{Fig3_Noise} shows the visual and quantitative results of amplitude (a) and phase (b). The results of Non-PSR lost target information due to insufficient pixels. Although the AP and WF algorithms obtain better resolutions, the super-resolution reconstruction further magnified measurement noise with degraded quality. Moreover, measurement noise resulted in a larger threshold of patterns number, they cannot utilize the advantage of fine modulation under even 100 patterns. In comparison, the reported algorithm can suppress background noise efficiently while maintaining fidelity. It also highlighted the strength of fine modulation with more details in the close-up.

\subsection{Experiments.}
We built a prototype of the reported lensless on-chip sensor, as shown in Fig. \ref{Fig1_system_architect} (d). We used a coverslip coated with $\sim$1$ \mu m$ polystyrene beads as the scattering layer and integrated it with an image detector with 1.67-$\mu m$ pixel size. The illumination source was a fiber-coupled laser diode (532 $nm$, 5 $mW$). The distance between the sample and the scattering layer was $\sim$1 $mm$. The diffraction distance between the scattering layer and the detector array was $\sim$0.5 $mm$. The reported prototype employed a unit magnification configuration with $\sim$50,000 Fresnel numbers (imaging area divided by 'd$\cdot$wavelength'). The unit magnification configuration makes the imaging FOV reach up to the entire sensor area (6.4 $mm$ by 4.6 $mm$).

\begin{figure*}[t!]
\centering
\includegraphics[width=\linewidth]{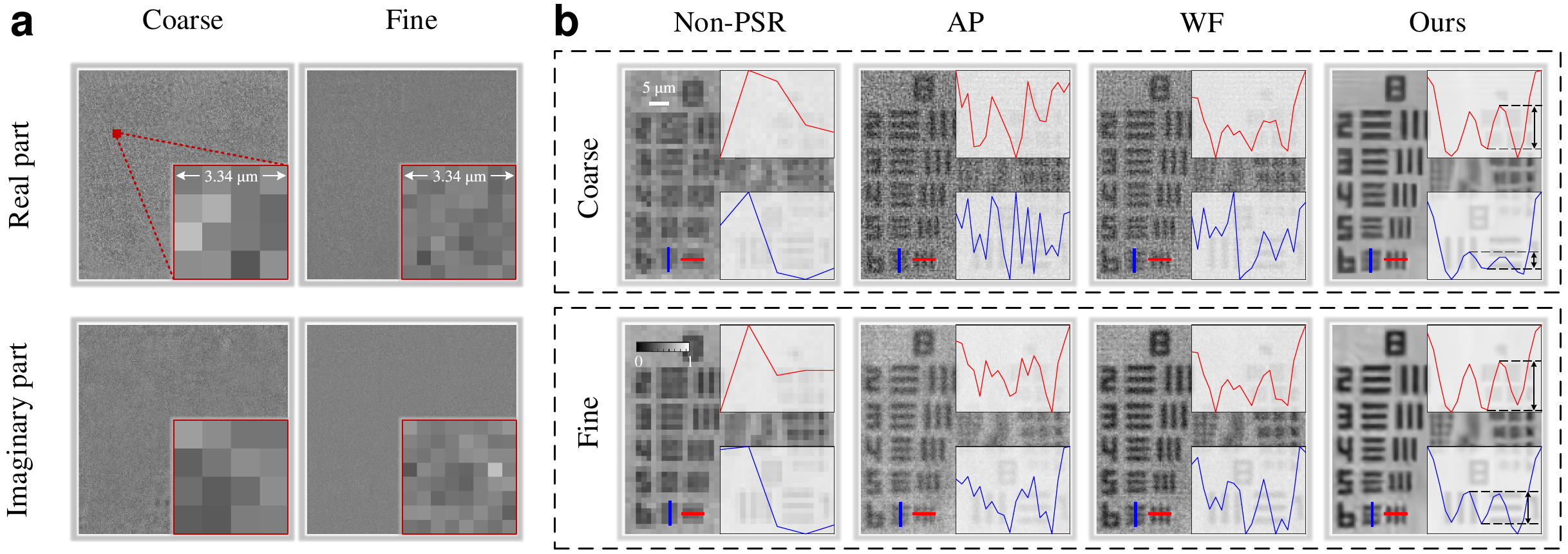}
\caption{Experimental results of the USAF resolution test chart. (a) is the calibrated complex-domain patterns of the scattering layer. (b) presents the amplitude results of the USAF resolution test chart. $\Delta_{M}$ of coarse and fine modulations were 0.835 $\mu m$ and 0.4175 $\mu m$, respectively.}
\label{Fig4_USAF}
\end{figure*}

Before imaging, we first calibrated the modulation pattern of the scattering layer (Supplementary Note 1) and obtained two modulation sizes of $\Delta_{M}=0.835$ $\mu m$ and $\Delta_{M}=0.4175$ $\mu m$. Then, we moved the integrated sensor to 225 different positions and captured corresponding intensity-only measurements for each sample. In PSR reconstruction, we set $\theta=4$ and used different modulation sizes to recover the high-resolution wavefront at the scattering layer plane. After decoding the scattering pattern, the wavefront was then propagated to the sample plane to generate the sample's wavefront.

Figure \ref{Fig4_USAF} (a) shows the calibrated complex-domain scattering layer. Figure \ref{Fig4_USAF} (b) presents the amplitude results of a USAF resolution test chart. The first row shows the results of coarse modulation ($\Delta_{M}=0.835$ $\mu m$) and the second row shows the results of fine modulation ($\Delta_{M}=0.4175$ $\mu m$). From the aspect of modulation sizes, the results of fine modulation obtain more smooth background and higher contrast (as presented in the cross sections). From the aspect of reconstruction algorithms, the Non-PSR had trouble recovering the feature of group 8. The AP and WF algorithms suffer from quality degradation due to the serious noise, which can only resolve the feature of group 8, element 5. In contrast, the reported method has a clearer background and higher resolution to resolve group 8, element 6. 

\begin{figure*}[t!]
\centering
\includegraphics[width=\linewidth]{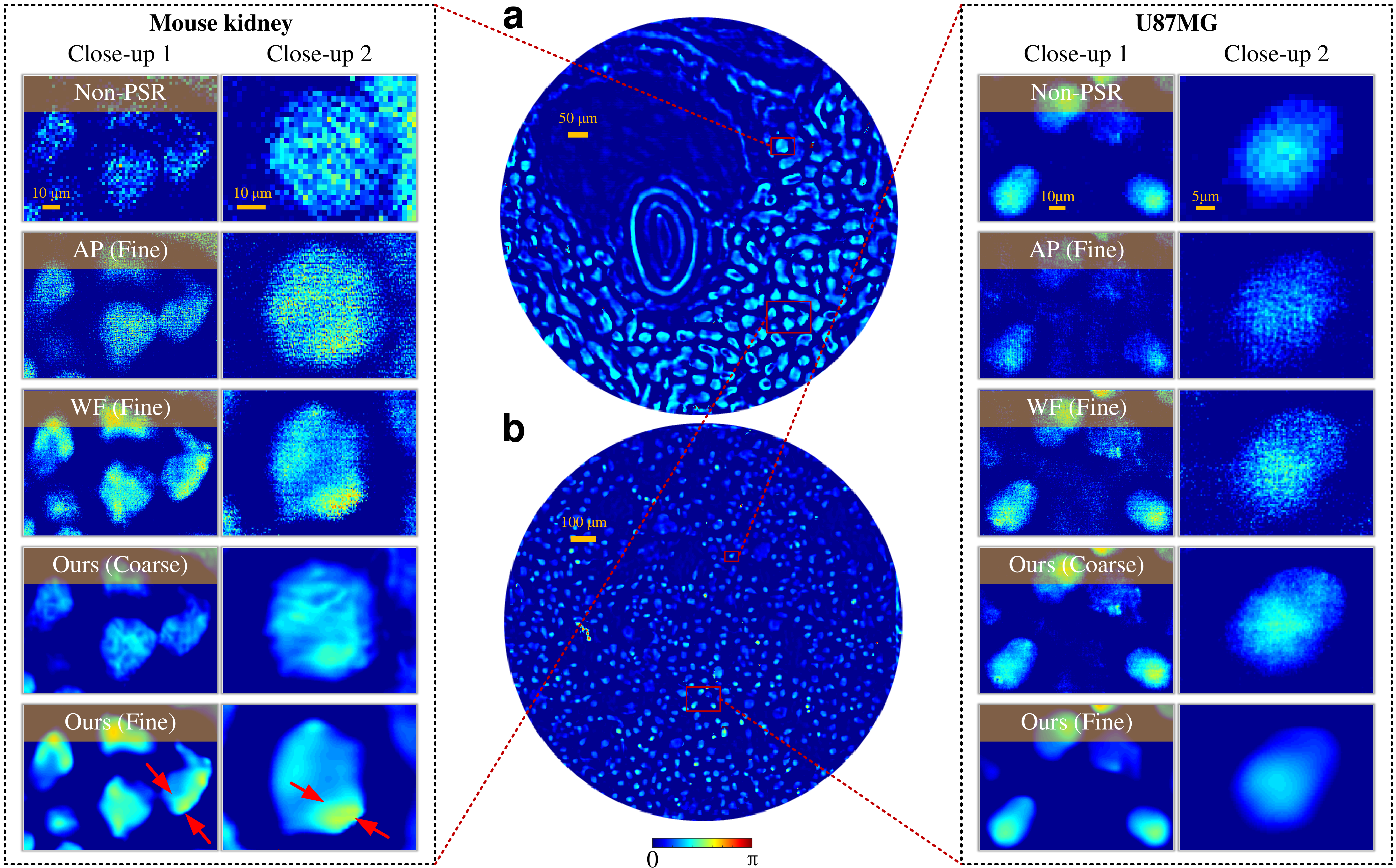}
\caption{Experimental results of unstained cell samples. (a) The phase images of the mouse kidney slide. (b) The phase images of the U87MG cells.}
\label{Fig5_Cells.pdf}
\end{figure*}

Figure \ref{Fig5_Cells.pdf} shows the recovered quantitative phase images of an unstained mouse kidney slide and U87MG cell culture. We can see that the results without PSR maintain ambiguous cellular outlines. The conventional PSR techniques can enhance resolution, but the image contrast is still unsatisfactory with disturbing noise and aberrations. In comparison, the reported technique produces high-fidelity results that effectively preserve fine details while attenuating unpleasant artifacts. Besides, the fine modulation can successfully recover high-density areas (as indicated by the arrows in close-ups), which indicate the potential cell overlap in three dimensions.

\begin{figure}[t!]
\centering
\includegraphics[width=0.5\linewidth]{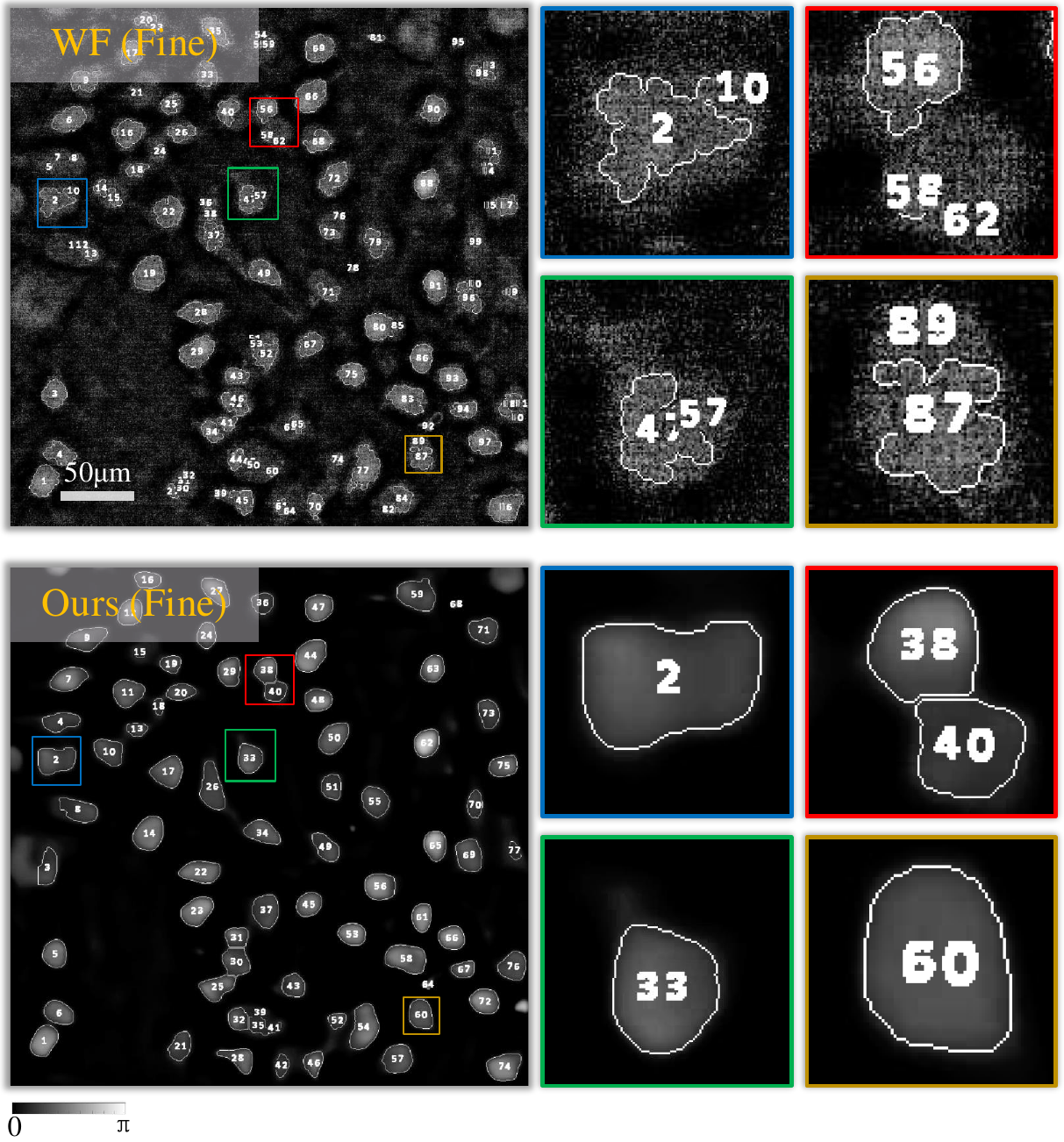}
\caption{Segmentation results of U87MG cells. We employed the watershed algorithm to segment the reconstructed images of different PSR techniques.} %The unpleasant reconstruction quality of the conventional PSR methods makes it hard for them to process the overlapping or connective cells, leading to incorrect segmentation or even omissive counting (see close-ups). In comparison, the segmentation result of DONet-PSR is more accurate.}
\label{fig:cell_segmentation}
\end{figure}

The state-of-the-art performance of the constructed sensor and the reported reconstruction algorithm benefits subsequent image analysis and intelligent processing. As a demonstration, we employed the watershed algorithm \cite{roerdink2000watershed} to implement cell segmentation and counting on the reconstructed phase images of U87MG cell culture. Specifically, we selected a $\sim$ 350 $\mu m$$^2$ square region and binarized the images to separate the U87MG cells from the background. After image filling, we employed watershed transform to perform cell segmentation. The marginal cells were removed to eliminate counting errors. Figure \ref{fig:cell_segmentation} presents the segmentation results of U87 cells. We can see that the WF algorithm with fine modulation cannot produce smooth edges and exist wrong segmentation. In comparison, the reported technique can improve the segmentation accuracy, producing high-fidelity results.
Table \ref{tab1:cell_counting} shows the cell counting results. For the selected $\sim$ 350 $\mu m$$^2$ region, there is an average of 72 cells via manual counting by three persons. We can see that the counting result of the reported algorithm with fine modulation (Ours-Fine) is most close to manual counting with only 6.9$\%$ counting error, which is eighth than the AP-Fine algorithm. In summary, the unpleasant reconstruction quality of the conventional PSR techniques makes automatic cell segmentation challenging. The reported techniques are able to produce a clear cell structure and provide an effective solution for rapid label-free cell segmentation and counting.

\begin{table}[t!]
\centering
\fontsize{13}{18}\selectfont  
\caption{The results of cell counting. 
There are 72 cells in the selected region via manual counting.}
\label{tab1:cell_counting}
\begin{tabular}{c|c|c}
\hline
Methods & Cell number &Error\\
%\midrule
\hline
AP-Fine   & 114 & 58.3$\%$           \\
WF-Fine    & 111          & 54.1$\%$ \\
Ours-Coarse     & 86         & 19.4$\%$ \\
Ours-Fine     & 77         & 6.9$\%$ \\
Ground Truth    & 72& ---    \\
%\bottomrule  
\hline 
\end{tabular}
\end{table}

\section{Conclusion and discussion}
In this work, we constructed a novel lensless on-chip sensor via scattering multiplexing and reported a robust pixel super-resolution (PSR) reconstruction algorithm. Benefiting from the high degree-of-freedom reconstruction of the scattering layer, we realize fine wavefront modulation at a sub-pixel size. The fine modulation can efficiently encode target information, improving the information throughput. Besides, the unit magnification configuration brought a large Fresnel number ($\sim$50,000), which can realize wide-field imaging. The reported robust PSR algorithm decomposed the reconstruction task into independent sub-problems via ADMM framework. The Wirtinger gradient solver and enhancing neural network solver were respectively derived to tackle the measurement formation and statistical prior regularization with high efficiency. Extensive simulations and experiments validated that the reported techniques provide a high-throughput solution for quantitative wavefront detection and subsequent image analysis.

The modulation characteristic of the scattering layer is related to its material and preparation technology. We can explore different scattering layers to further optimize target encoding and information throughput. Besides, recent advanced in optical metamaterial \cite{cui2014coding,dorrah2022tunable} and whole imaging flow learning \cite{zhang2022end} techniques allow the joint optimization of optical component and imaging algorithm, which may be an important research field for the reported techniques.

The reported reconstruction algorithm can be also extended in the future. First, it involves multiple parameters that are currently adjusted manually. We can introduce the reinforcement learning technique to automatically adjust parameters \cite{wei2020tuning}. Second, we can introduce the parallel computing technique to effectively improve running efficiency \cite{sharma2009matlab}. Third, since the reported technique is to increase resolution, it is interesting to study introducing the super-resolution network \cite{zhang2019deep,wang2020deep} as the enhancement operator, which may further improve resolution.


\begin{suppinfo}
Supporting Information Available: the details of scattering layer calibration and the derivation of Wirtinger gradient. 
This material is available free of charge via the Internet at http://pubs.acs.org
\end{suppinfo}

\section{Author Information}
\subsection{Author Contributions
}
X.C. and L.B. conceived the idea and supervised the project. X.C., S.J. and Y.H. performed the algorithm study and data analysis. S.J. and G.Z. prepared the integrated sensor and collected data. All the authors contributed to writing and revising the manuscript, and participated in discussions during the project.
\subsection{Funding}
This work was supported by the National Natural Science Foundation of China (Nos. 61827901, 61991451, 62131003), BIT Research and Innovation Promoting Project (No. 2022YCXZ006).
\subsection{Notes}
The authors declare no competing financial interest.

%\begin{acknowledgement}
%The authors thank Yunhui Gao and Liangcai Cao (Tsinghua University) for helpful discussion in theoretical derivation.
%\end{acknowledgement}

\bibliography{ref}

\end{document}